\documentclass[%
 reprint,
 superscriptaddress,
 amsmath,amssymb,
prb,
citeautoscript
]{revtex4-1}
\usepackage[normalem]{ulem}
\usepackage{natbib}
\usepackage{xcolor}
\usepackage{graphicx}
\usepackage{amsmath}
\usepackage{physics}
\usepackage[colorlinks=true,allcolors=blue]{hyperref}%
\DeclareUnicodeCharacter{2212}{-}

\usepackage{xspace}

\newcommand{\CCO}{$\mathrm{CaCoO}_2$\xspace}
\newcommand{\CCOB}{$\mathrm{CaCoO}_{2.5}$\xspace}

\renewcommand{\sout}[1]{\unskip}
\newcommand{\figref}[2]{\hyperref[#1]{Figure~\ref{#1}(#2)}}
\newcommand{\figureref}[2]{\hyperref[#1]{Figure~\ref{#1}(#2)}}

\usepackage{lineno}
\usepackage{nameref}
\usepackage{hyperref}
\hypersetup{colorlinks,allcolors=blue}

\begin{document}
% \setcounter{secnumdepth}{0}

%Title of paper
\title{Orbital inversion and emergent lattice dynamics in infinite layer CaCoO$_\mathbf{2}$ } 
\date{\today}
\author{Daniel~Jost}
\email{daniel.jost@stanford.edu}
\affiliation{Stanford Institute for Materials and Energy Sciences, SLAC National Accelerator Laboratory, Menlo Park, CA, USA}

\author{Eder~G.~Lomeli}
\affiliation{Stanford Institute for Materials and Energy Sciences, SLAC National Accelerator Laboratory, Menlo Park, CA, USA}
\affiliation{Department of Materials Science and Engineering, Stanford University, Stanford, CA, USA}

\author{Woo~Jin~Kim}
\affiliation{Stanford Institute for Materials and Energy Sciences, SLAC National Accelerator Laboratory, Menlo Park, CA, USA}

\author{Emily~M.~Been}
\affiliation{Stanford Institute for Materials and Energy Sciences, SLAC National Accelerator Laboratory, Menlo Park, CA, USA}

\author{Matteo~Rossi}
\affiliation{Stanford Institute for Materials and Energy Sciences, SLAC National Accelerator Laboratory, Menlo Park, CA, USA}

\author{Stefano~Agrestini}
\affiliation{Diamond Light Source, Harwell Campus, Didcot OX11 0DE, UK.}

\author{Kejin~Zhou}
\affiliation{Diamond Light Source, Harwell Campus, Didcot OX11 0DE, UK.}

\author{Chunjing~Jia}
\affiliation{Stanford Institute for Materials and Energy Sciences, SLAC National Accelerator Laboratory, Menlo Park, CA, USA}

\author{Brian~Moritz}
\affiliation{Stanford Institute for Materials and Energy Sciences, SLAC National Accelerator Laboratory, Menlo Park, CA, USA}

\author{Zhi-Xun~Shen}
\affiliation{Stanford Institute for Materials and Energy Sciences, SLAC National Accelerator Laboratory, Menlo Park, CA, USA}
\affiliation{Geballe Laboratory for Advanced Materials, Stanford University, Stanford, CA, USA}
\affiliation{Department of Applied Physics, Stanford University, Stanford, CA, USA}
\affiliation{Department of Physics, Stanford University, Stanford, CA, USA}

\author{Harold~Y.~Hwang}
\affiliation{Stanford Institute for Materials and Energy Sciences, SLAC National Accelerator Laboratory, Menlo Park, CA, USA}
\affiliation{Geballe Laboratory for Advanced Materials, Stanford University, Stanford, CA, USA}
\affiliation{Department of Applied Physics, Stanford University, Stanford, CA, USA}

\author{Thomas~P.~Devereaux}
\email{tpd@stanford.edu}
\affiliation{Stanford Institute for Materials and Energy Sciences, SLAC National Accelerator Laboratory, Menlo Park, CA, USA}
\affiliation{Department of Materials Science and Engineering, Stanford University, Stanford, CA, USA}
\affiliation{Geballe Laboratory for Advanced Materials, Stanford University, Stanford, CA, USA}

\author{Wei-Sheng~Lee}
\email{leews@stanford.edu}
\affiliation{Stanford Institute for Materials and Energy Sciences, SLAC National Accelerator Laboratory, Menlo Park, CA, USA}

\date{\today}

%For the arXiv post
\begin{abstract}
The layered cobaltate \CCO exhibits a unique herringbone-like structure. Serving as a potential prototype for a new class of complex lattice patterns, we study the properties of \CCO using X-ray absorption spectroscopy (XAS) and resonant inelastic X-ray scattering (RIXS). Our results reveal a significant inter-plane hybridization between the Ca $4s-$ and Co $3d-$orbitals, leading to an inversion of the textbook orbital occupation of a square planar geometry. Further, our RIXS data reveal a strong low energy mode, with anomalous intensity modulations as a function of momentum transfer close to a quasi-static response suggestive of electronic and/or orbital ordering. These findings indicate that the newly discovered herringbone structure exhibited in \CCO may serve as a promising laboratory for the design of materials having strong electronic, orbital and lattice correlations. 
\end{abstract}

\maketitle

\section{Introduction}
Advancements in materials science have underscored how structural motifs dictate the properties and functionalities of complex materials~\cite{Takada:2003,Roger:2007, Senn:2012,Green:2014, Stoerzinger:2015, Yu:2017, Burch:2018, Li:2019, Xu:2020, Wang:2023}. Here, we examine the newly discovered infinite layer material CaCoO$_2$ which features a unique super-structure~\cite{Kim:2023}.
Using X-ray absorption and resonant inelastic X-ray scattering~\cite{Ament:2011:RMP}, we uncover an inversion of the Co $\textit{3d}-$orbital occupations resulting from a significant inter-plane hybridization between Ca $\textit{4s}-$ and Co $\textit{3d}-$ orbitals. Additionally, we find a strong lattice vibrational mode having an intensity modulation near a structurally forbidden satellite peak of potentially electronic origin. This suggests the presence of strong electron-lattice correlations which may aid in stabilizing the \AA ngstrom-scale lattice distortions seen in CaCoO$_2$. Our work establishes this infinite-layer compound as a novel platform for the investigation of orbitally engineered systems inhabited by strong correlation effects.

\begin{figure*}
    \centering
    \includegraphics[width = 180mm]{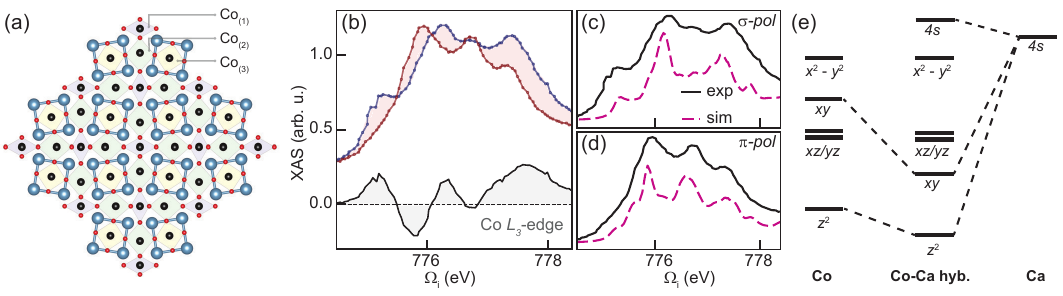}
    \caption{\textbf{Crystal structure of CaCoO$_\mathbf{2}$.} (a)~Top view of the infinite-layer cobaltate \CCO showing Ca atoms in blue, Co atoms in black and O atoms in red. Three distinct cobalt sites are identified as two distorted (blue Co$_{(1)}$ and green Co$_{2}$) and a rotated, undistorted site (yellow Co$_{(3)}$). (b)~Experimental X-ray absorption (XAS) results taken in total electron yield (TEY) for photon polarizations perpendicular ($\sigma$, blue) and parallel ($\pi$, red) to the scattering plane, at an incident angle of $\theta_\mathrm{in}=20^\circ$. The black curve corresponds to the linear dichroism $I_\mathrm{dic} = I_\sigma - I_\pi$, i.e. the difference of the signal in the two polarization channels. (c),(d)~Experimental data reproduced from panel (b) along with the dashed magenta lines which correspond to multiplet simulations using exact diagonalization. (e)~Energy ordering of the Co $3d-$orbitals for square planar geometry (left). Hybridization of the Ca $4s-$orbitals pushes the $3d_{xy}-$ and $3d_{z^2}-$orbitals to lower energy, leading to finite hole content in the $3d_{xz,yz}-$orbitals. $\Omega_\mathrm{i}$ is the incident photon energy given in electron-Volts (eV).}
    \label{fig:Fig1_XAS}
\end{figure*}

% %%%%%%%%%%%%%%%%%%%%%%%%%%%%%%%%%%%%%%%%%%%%
% \section*{Results}
% \label{sec:results}
Among infinite layer transition metal oxides which are known for hosting a variety of complex quantum phenomena~\cite{Siegrist:1988, Takada:2003, Tsujimoto:2007, Kawakami:2009, Li:2019, Kim:2023} \CCO stands out due to its unique crystal structure~\cite{Kim:2023}. This structure emerges upon reducing the Brownmillerite parent compound \CCOB via removal of its apical oxygen, in a similar fashion as the nickelate superconductors~\cite{Li:2019}. 
In contrast to its Ni-based siblings, however, \CCO does not collapse into a simple square-planar geometry but exhibits large displacements in both the Co-O bonds as well as the Ca layer [\figref{fig:Fig1_XAS}{a}]. This leads to three distinct Co sites, each experiencing different effective ligand fields, which modify the $\mathrm{Co^{2+}}$ electronic configuration with 7 electrons distributed in the $3d-$orbitals~\cite{Kim:2023},  resulting in three types of locally distorted CoO$_2$ plaquettes. With anomalously large, \AA-sized CoO$_2$ distortions, the pseudo-cubic unit cell reconstructs into an enlarged and geometrically frustrated $2\sqrt{2}\times 2\sqrt{2} \times 1$ superstructure. While DFT + \textit{U} calculations reproduce the overall crystal symmetry~\cite{Kim:2023}, they fail to match the values of the lattice distortions quantitatively. The experimental magnitude of these distortions alongside the quantitative disagreement with \textit{ab initio} theoretical results indicate that additional factors, for instance strong correlation effects both between charge carriers and with the lattice, might be at play.

\section{Methods}
\label{sec:methods}

\subsection{Sample synthesis}
Pulsed laser deposition was used to grow a $20\,\mathrm{nm}$ thin film of the parent Brownmillerite compound \CCOB on a SrTiO$_3$ (001) substrate. The films were capped with five unit cells of SrTiO$_3$ layers to prevent potential degradation during the reduction process. The precursor \CCOB films were then reduced to the infinite layer \CCO system through topotactic reduction~\cite{Kim:2023}.

\subsection{X-ray scattering}
X-ray Absorption Spectroscopy (XAS) and Resonant Inelastic X-ray Scattering (RIXS) measurements were performed at beamline i21 of the Diamond Light Source, United Kingdom. For the incident energy maps, the photon energy was tuned across the Co $L_3-$edge as measured from XAS. The incident photon polarization was set either parallel ($\pi$) or perpendicular ($\sigma$) to the scattering plane. The combined energy resolution was $36\,\mathrm{meV}$. The momentum dependent measurements were conducted along the (h,0), (h,k) and (-0.05,-0.05, l) directions with units given in reciprocal lattice units (r.l.u.) of the pseudocubic unit cell, having lattice parameters $a=b=3.9\,\mathrm{\AA}$, $c=3.27\,\mathrm{\AA}$ with conversions $(2\pi/a,2\pi/b,2\pi/c)$. The scattering angle was fixed to $154^\circ$ for the measurements along (h,0) and (h,k) and varied for measurements along the (-0.05,-0.05,l) direction. 

\subsection{Exact Diagonalization of Charge Transfer and Hybridization Full Atomic Multiplet (CTHFAM)}

To model the various local electronic environment of Co in this material, we exactly diagonalize a full atomic multiplet including charge transfer and hybridization effects of a Co transition metal center (5 d orbitals), 2 oxygen ligands (3 p orbitals each) and one Ca ligand (1 s orbital). The single site is centered at (\begin{math}\pi/2,\pi/2\end{math}) in momentum. From the formal valence of $\mathrm{CaCoO_2}$, $\mathrm{Co^{+2}}$ and $\mathrm{Ca^{+2}}$  each contribute 3 and 2 holes, respectively. From this configurational space, the interactions are represented by the following Hamiltonian:
\begin{align}
    \hat H &= 
     \frac{1}{2}\sum_{i,\sigma,\sigma'} \sum_{\mu,\nu,\mu',\nu'} U_{\mu,\nu,\mu',\nu'} \hat c^\dagger_{i,\mu,\sigma}\hat c^\dagger_{i,\nu,\sigma'}\hat c_{i,\mu',\sigma'}\hat c_{i,\nu',\sigma} \nonumber\\ 
     & + \sum_{i,j,\sigma}\sum_{\mu,\nu} t_{i,j}^{\mu,\nu} \hat c^\dagger_{i,\mu,\sigma}\hat c_{j,\nu,\sigma} \nonumber + \sum_{i,\mu,\nu,\sigma}V_{CEF}(\mu,\nu)c^\dagger_{i,\mu,\sigma}\hat c_{i,\nu,\sigma}\nonumber \\
& + \frac{1}{2}\sum_{i,\sigma,\sigma'} \sum_{\mu,\nu,\mu',\nu'} U_{\mu,\nu,\mu',\nu'} \hat c^\dagger_{i,\mu,\sigma}\hat d^\dagger_{i,\nu,\sigma'}\hat c_{i,\mu',\sigma'}\hat d_{i,\nu',\sigma} \nonumber\\
& - \sum_{i,\sigma,\sigma'} \sum_{\mu,\nu} \lambda_{\mu,\nu}^{\sigma,\sigma'} \hat d^\dagger_{i,\mu,\sigma} \hat d_{i,\nu,\sigma'} + \sum_{i}\Delta_{i}n_{i}
\end{align}
Where \textit{i, j} refer to the different atomic sites, $\mu$, $\nu$ refer to different sets of \textit{l, m} quantum numbers, and $\sigma$ refers to spin. The first term includes a Hubbard-like U term for the coulomb interaction where all Co sites exhibit a \begin{math}U = 5.4\end{math} and \begin{math}J_H = 0.9\end{math}, typical values for TM oxides. The second term includes a \textit{t} hopping element between different atomic sites and their orbitals. The third term includes an octahedral crystal field for the d-orbtials in the metal atom, the fourth term is the core-valence coulumb interaction, the fifth term is the spin-orbit coupling $\lambda$ at the core and the last term is the charge transfer energy $\Delta$ at each atomic site. For a list of the parameters used to model each of the three different cobalt environments, see supplementary table X. The multi-particle eigenstates for a Hamiltonian of an \textit{N} hole cluster and one for a \textit{N-1} hole cluster with a core hole serve as the initial (\textit{i}), intermediate ($\nu$), and final (\textit{f}) states for the calculation of XAS by Fermi's golden rule:

\begin{align}
&\kappa_{e_i, k_i}(\omega)=\nonumber \\
&\frac{1}{\pi Z} \sum_{i,\nu} e^{-\beta E_i} \mid \langle\nu\mid \hat D_{k_i}({e_i})\mid i \rangle\mid^2 \delta(\omega-(E_\nu-E_i))
\end{align}

And for RIXS using the Kramers-Heisenberg representation:

\begin{align}
&R( e_i,e_f,k_i,k_f,\omega_i,\omega_f)=\nonumber\\ &\frac{1}{\pi Z} \sum_{i,f} e^{-\beta E_i} \left\vert \sum_\nu \frac{\langle f \mid \hat D_{k_f}^*(e_f)\mid\nu\rangle \langle\nu\mid \hat D_{k_i}({e_i})\mid i\rangle}{\omega_i-(E_\nu-E_i)-i \Gamma}\right\vert^2 \nonumber \\
&\delta(\Omega-(E_f-E_i))
\end{align}

Where \begin{math}E_{i,\nu,f}\end{math} refers to the eigenenergy and \begin{math} D_{k_i}({ e_i})\end{math} is the dipole operator for a photon of frequency $\omega$, momentum \textit{k} and polarization \textit{e}, and $\Omega=\omega_i-\omega_f$. A core-hole lifetime $\Gamma$ broadening of 0.2 eV was used for the plotting of theoretical XAS/RIXS maps. A global energy shift for each site was also added to all calculated spectral features for ease of comparison with experiment. The final simulated spectra was constructed using 1/2/1 ratios for the distinct Co sites individual simulations, based on how many of each site is found on the unit cell of the material. An average over X and Y incoming photon polarizations were used for the simulation of the fully in-plane \begin{math}\sigma\end{math}-polarization experiments, and the mostly out-of-plane \begin{math}\pi\end{math}-polarization experiments were simulated with 87\% contribution of Z photon polarization and 93\% in-plane contribution for XAS and with 7\% contribution of Z photon polarization and 13\% in-plane contribution for RIXS, based on the experimental set up and incident angle for each measurement.
%%%%%%%%%%%%%%%%%%%%%%%%%%%%%%%%%%%%%%%%%%%%

\section{Results and Discussion}

The electronic structure can be inferred from the rich multiplet splitting of the Co $L_3-$edge XAS [\figref{fig:Fig1_XAS}{b}], providing a basis for formulating an effective low energy model. A strong dependence of the incident photon polarization can be seen when tuned either parallel ($\pi$) or perpendicular ($\sigma$) to the scattering plane.
The two polarization channels highlight distinct multiplet structures at associated incident energies which is more apparent in the difference spectrum shown in  \figref{fig:Fig1_XAS}{b}. This is in contrast to, for instance, cubic CoO~\cite{Regan:2001, Isotim:2015} in which the multiplet structure is isotropic with respect to the incident linear photon polarizations. On a qualitative level, the strong dichroism indicates that the orbital structure of \CCO hosts empty states having $z-$components, since the $\sigma$ and $\pi$ photon polarizations in our experimental geometry highlight projected weights of the $3d-$orbitals in- and out-of-the CoO$_2$ plane, respectively. In \CCO, however, we find the experimentally observed multiplet splitting cannot be reproduced via exact diagonalization (ED) calculations using the typical orbital energetic sequence of a CoO$_6$ tetrahedron [see \figref{fig:Fig1_XAS}{e} for Co and~\autoref{fig:ED_fails_suppl}], signaling an abnormal electronic structure of the Co ion. 

In the absence of any other apical atomic species that could cause such a change of the orbital occupations, we conjecture that strong overlap between Ca $4s-$ and Co $3d-$orbitals, facilitated by an atypically shorter c-axis lattice constant, would lead to a significant orbital sequence reordering resulting in the observed dichroism. Considering the phase factors of the Co $3d-$orbitals, the strongest overlap with the $4s-$orbital is to be expected with the $d_{xy}-$ and $d_{z^2}-$orbitals, leading to an effective orbital inversion of the $d_{xz,yz}-$ with the $d_{xy}-$orbital occupations. We find validation thereof in a comparison of our experimental results with ED simulations in which we account explicitly for the orbital inversion on all Co sites, as well as the Ca $4s-$hybridization [\figref{fig:Fig1_XAS}{c,d}]. This combined qualitative and quantitative assessment reveals the local electronic structure depicted in \figref{fig:Fig1_XAS}{e} with the holes distributed mainly in the $d_{x^2-y^2}-$ and half-filled $d_{xz,yz}-$orbitals, alongside the almost entirely empty $4s-$orbitals. This picture derived from the XAS and ED simulations reveals that the distortion may be at least stabilized and possibly driven by a strong interlayer hybridization effect.   

\begin{figure*}
    \centering
    \includegraphics[width = 172mm]{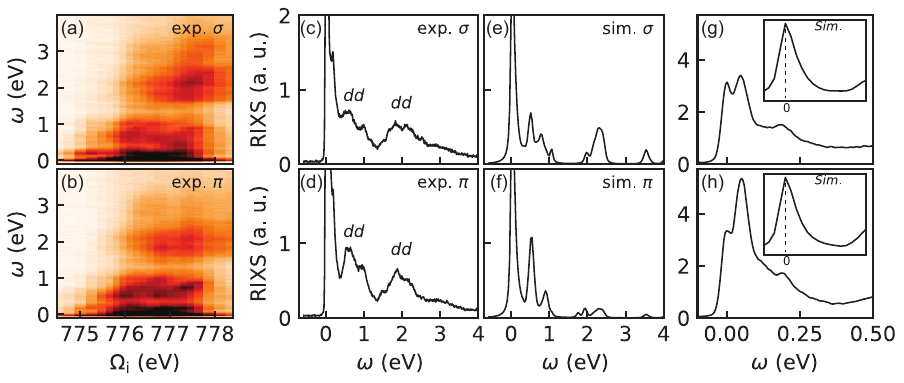}
    \caption{\textbf{RIXS signal across the \textit{L}$_\mathbf{3}$-edge of CaCoO$_\textbf{2}$}. (a),(b)~Experimental results for $\sigma$- and $\pi-$polarization. (c), (d)~Line-cuts of the experimental color maps at $\Omega_\mathrm{i} =776.85\,\mathrm{eV}$. (e),(f) RIXS plots of the ED calculations using the electronic model presented in \autoref{fig:Fig1_XAS} to simulate the XAS response. (g),(h)~Same data as in (c),(d) zoomed to smaller energy transfer showing low energy features not captured by the simulations (insets). The energy transfer is given as $\omega$ in electron-Volts (eV).}
    \label{fig:Fig2_dd}
\end{figure*}

These results are also consistent with the orbital excitation spectrum seen in RIXS: \autoref{fig:Fig2_dd} illustrates the RIXS response for the two different polarization channels $\sigma$ and $\pi$ with several clusters of excitations visible. We find peak structures in two bands from 300 meV to up to 1.2\,eV and again from 1.2 to 2.8\,eV followed by a broad response extending to 4.0\,eV [\figref{fig:Fig2_dd}{c,d}]. These excitations can be identified as intra-atomic $dd-$excitations and are consistent with ED calculations [\figref{fig:Fig2_dd}{e,f}] obtained from the electronic structure model described in \autoref{fig:Fig1_XAS}. Interestingly, zooming in closer to a $300\,\mathrm{meV}$ energy scale [\figref{fig:Fig2_dd}{g,h}], there is a strong feature on a 40\,meV scale, followed by a broad band of excitations. These low energy excitations are not captured by the ED simulations [insets of \figref{fig:Fig2_dd}{g,h}], and thus cannot be attributed to $dd-$ excitations.

\begin{figure*}
    \centering
    \includegraphics[width = 120 mm]{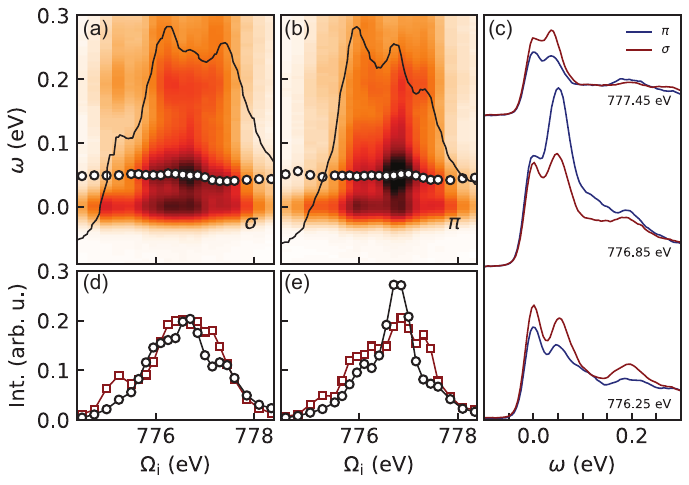}
    \caption{\textbf{Incident photon energy dependence of the low energy features.} (a)~$\sigma-$polarization. (b)~$\pi$-polarization. Filled markers in (a) and (b) correspond to the results of a fitting procedure. (c)~Linecuts at incident energies $\Omega_\mathrm{i}$ as indicated. A strong feature off of the elastic line is visible, followed by broad features at higher energy. (d),(e)~Intensity distribution of the mode across the $L_3-$edge for $\sigma-$ and $\pi-$polarization (black markers) and of the higher energy feature (red markers) centered at $\sim 180\,\mathrm{meV}$. The mode intensity peaks at an energy corresponding to the XAS maximum of the square planar site for $\pi-$polarization. The intensities correspond to the results of the fitting procedure seen in \autoref{fig:incmapsfits_LV} and \autoref{fig:incmapsfits_LH}. }
    \label{fig:Fig3_mode}
\end{figure*}

We first focus our discussion on the sharp low energy mode with an energy scale of 40 meV. At first glance, given the sharpness and bright intensity, this feature may be reminiscent of an exciton, but the transport gap of $\sim 0.3\,\mathrm{eV}$~\cite{Kim:2023} immediately rules out this option, as excitons are present only at distinct energies in the RIXS signal~\cite{Kang:2020}, whereas the feature is seen throughout the Co $L_3-$absorption edge, albeit with differing intensities. Considering that the energy scale is consistent with typical oxygen phonons in other 3$d-$ transition metal oxides, the most likely scenario is that the mode represents lattice vibrations with oxygen character. \figref{fig:Fig3_mode}{a,b} displays the RIXS intensity map at low energy as a function of the incident photon energy. The 40 meV mode has a strong resonance across the Co L$_3$-edge for both polarization channels, with a clear difference in the intensity distribution [\figref{fig:Fig3_mode}{c}]: the mode sharpens for $\pi-$polarization, and shows an intensity maximum at a specific incident energy  of $776.75\,\mathrm{eV}$ [\figref{fig:Fig3_mode}{d,e}]. The resonant profile of the mode has a maximum at an energy where the partial XAS associated with the undistorted Co site also contributes more significantly [see \textit{cf.} \autoref{fig:ED_indvidual}], indicating locally stronger coupling of this mode to that Co-site. 

As depicted in \figref{fig:Fig3_mode}{d,e}, the broad-band excitations observed within the 100-250 meV energy range exhibit resonant profiles that closely reflect the XAS spectrum and differ markedly from those of the 40 meV mode. This disparity in resonant behavior strongly suggests that these excitations are unrelated to the 40 meV mode. We hypothesize that these broad features may originate from polaronic interactions with other lattice vibrational modes~\cite{Jost:2024}. Furthermore, excitonic states cannot be excluded as potential contributors to these excitations, given that the particle-hole RIXS final state could exhibit extended lifetime due to an energy scale on the order of the band gap deduced from transport measurements. A comprehensive discussion about the nature of these excitations necessitates further investigation beyond the purview of this study. Accordingly, the remainder of this work will concentrate on a detailed analysis of the sharp 40 meV mode.

\begin{figure*}
    \centering
        \includegraphics[width = 120mm]{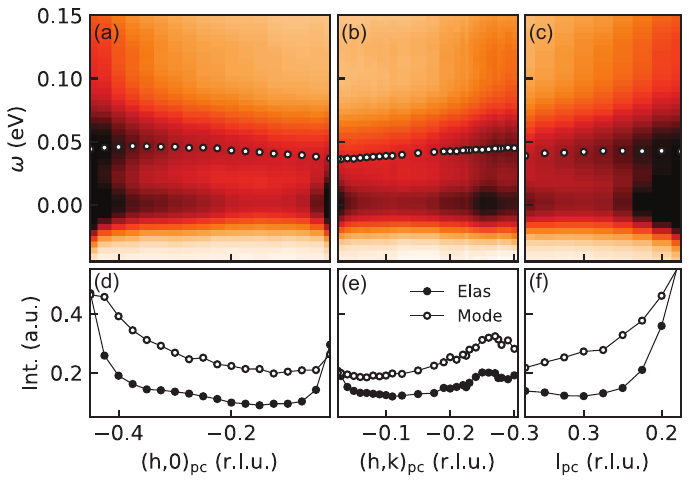}
    \caption{\textbf{Momentum dependence of the low energy mode.} (a)-(c)~Momentum dependence of the low energy mode along the (h,0), (h,k) and l-direction of the pseudo-cubic unit cell in r.l.u. (reciprocal lattice units). The measurements along the l-direction were taken at fixed in-plane momentum (-0.05, -0.05) r.l.u. (d)-(f)~Intensity distribution of the elastic line and the mode.}
    \label{fig:Fig4_JT_mode}
\end{figure*}

The momentum dependence of the sharp mode along different high symmetry directions in the Brillouin zone, depicted in \autoref{fig:Fig4_JT_mode}, corroborates its assignment as a phonon mode and rules out a magnetic origin, whose dispersion should eminate from $\omega = 0$ at zone center, before reaching a maximum at some finite momentum, and terminate at $\omega = 0$ approaching the magnetic ordering vector. Here, along the pseudo-cubic (h,0)-direction [\figref{fig:Fig4_JT_mode}{a}], the mode disperses from close to zone center at $\sim 35\,\mathrm{meV}$, raising to 45\,meV before slightly bending downwards. Similarly, the response along the (h,k)-direction shows a small slope [\figref{fig:Fig4_JT_mode}{b}], as does the out-of-plane component [\figref{fig:Fig4_JT_mode}{c}], indicating that the mode corresponds to an optical phonon branch. 

The scattering intensity of this phonon branch features a strong increase in intensity close to $\mathbf{q}_\mathrm{O} = (1/4,1/4)$ r.l.u., where the quasi-elastic intensity also exhibits a peak-like feature [\figref{fig:Fig4_JT_mode}{b,e}]. While $\mathbf{q}_\mathrm{O}$ reflects the periodicity of the large superstructure of \CCO (\figref{fig:Fig1_XAS}{a}), scattering form factor calculations indicate a forbidden superstructure peak that should have zero scattering intensity at $(1/4, 1/4, 0)$~\cite{Kim:2023}. Rather than a lattice Bragg reflection, the peak likely is associated with electronic order with a characteristic wave-vector $\mathbf{q}_\mathrm{O}$. While the origin of this order manifested in the quasi-elastic intensity profile remains an open question, the fact that the phonon intensity shows a similar enhanced at $\mathbf{q}_\mathrm{O} = (1/4,1/4) r.l.u$, indicates that the mode may be strongly coupled to this order. This is quite reminiscent of the phonon intensity anomaly observed in some cuprates~\cite{Lee:2021, Arpaia:2023}, with an enhanced intensity near the CO wave-vector. 

The phonon intensity in RIXS reflects the charge-lattice coupling strength~\cite{Ament:2011_EPC,Devereaux:2016,Braicovich:2020}, such that the phonon intensity modulation at $\mathbf{q}_\mathrm{O}$ may be indicative of a spatially dependent coupling which reinforces the structural distortion. In other words, free energy is reduced by intertwined orbital, charge, and lattice contributions -- strong hybridization, formation of some electronic order, and enhanced charge-lattice coupling, contributing to the large distortions seen in \CCO.  

As previously discussed in \figref{fig:Fig3_mode}{d,e} this phonon, which couples to electronic-order, couples strongly to the un-distorted Co site among the three environments, which itself is rotated concomitantly with the Ca cage. We suspect that the orbital inversion facilitated by the strong Ca hybridization with this Co site may play an important role in the formation and properties of the electronic order. Such a lattice-enforced orbital inversion in \CCO is quite unique, absent in infinite-layer nickelates and cuprates, and rare even among other transition metal oxides. These aspect calls for a more thorough investigation of the \CCO compound. Note that previous DFT calculations may underestimate the electron-lattice coupling, as the effect of orbital inversion may not be fully accounted for with the prediction of such a small distortion~\cite{Kim:2023}. Our results place additional constraints on the formulation of an effective low-energy model for the \CCO system.

\section{Conclusion}

In summary, the herringbone structure of \CCO represents a compelling new platform for manipulating and studying exotic material properties. The orbital inversion facilitated by the Ca hybridization could imprint distinct features on the electronic structure, potentially paving the way towards engineered topology. Moreover the spectroscopic fingerprints of strong electronic and lattice correlations raises questions about the effects of doping on the transport properties of \CCO, in particular, whether the orbital inversion, electronic order, and the super-cell persist with doping and whether superconductivity can emerge, like in its nickelate and cuprate cousins. These possibilities underscore the promising outlook for \CCO in future material science research.   \\

\paragraph*{Author contributions} 
D.J. and W.S.L. conceived the experiment and analyzed the data. W.J.K. synthesized the samples. D.J., M.R., S.A., K.Z., and W.S.L. performed the experiments. E.G.L., E.M.B., C.J., B.M., and T.P.D. performed the theory work. Z.X.S., H.Y.H., T.P.D., and W.S.L. supervised the project. All authors contributed to the interpretation. D.J. and W.S.L. wrote the manuscript with input from all authors. 

\paragraph*{Competing financial interests:} The authors declare that there are no competing financial interests.

\textbf{Data availability statement:} Data are available from the corresponding authors upon reasonable request.

\begin{acknowledgments}
The work was supported by the U.S. Department of Energy, Office of Basic Energy Sciences, Materials Sciences and Engineering Division. XAS and RIXS measurements were performed at beamline I21, Diamond Light Source (UK). D.J. gratefully acknowledges funding by the Alexander-von-Humboldt foundation. 

Aspects of the materials development were supported by the Gordon and Betty Moore Foundation’s Emergent Phenomena in Quantum Systems Initiative (grant no. GBMF9072).

Computational work was performed on the Sherlock cluster at Stanford University and on resources of the National Energy Research Scientific Computing Center (NERSC), a Department of Energy Office of Science User Facility, using NERSC award BES-ERCAP0027203.
\end{acknowledgments}

% \newpage
%%%%%%%%%%%%%%%%%%%%%%%%%%%%%%%%%%%%%%%%%%%

\bibliography{cobaltates}

% %\linespread{2}

\clearpage
\newpage 
\onecolumngrid

\setcounter{page}{1}

\renewcommand\thefigure{S\arabic{figure}}
\renewcommand\thepage{S\arabic{page}}
\renewcommand\thesection{S\arabic{section}}
\setcounter{figure}{0}

\section*{Supplementary Material}

\begin{figure*}[h!]
    \centering
    \includegraphics[width = 85mm]{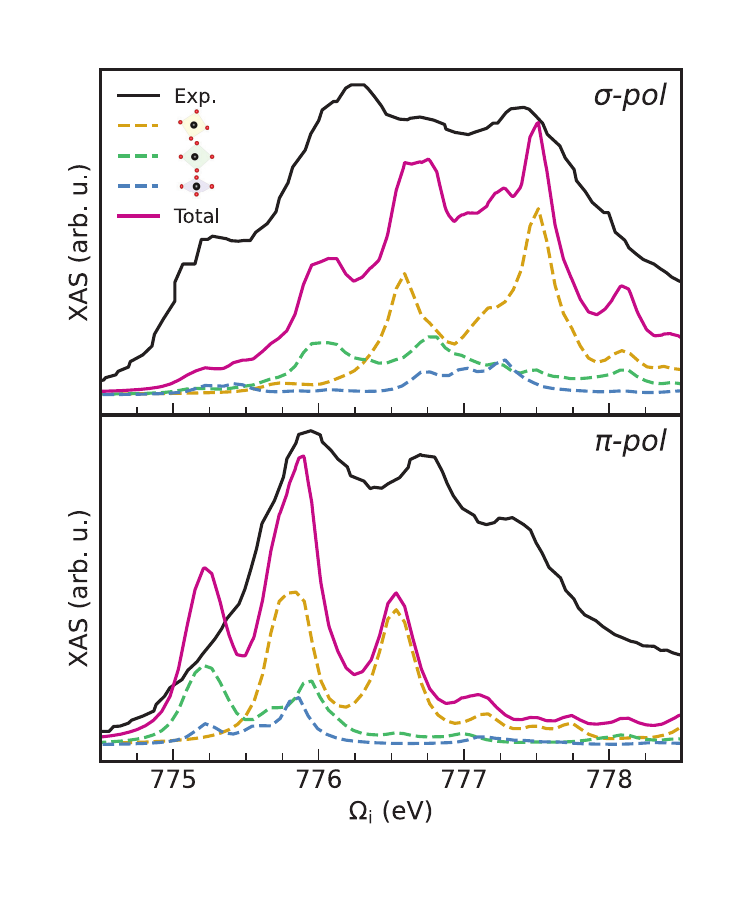}
    \caption{\textbf{Multiplet calculations with square planar crystal field.} Individual contributions of each cobalt environment to the overall theoretical result are depicted in colored dashed lines for $\sigma$ and $\pi$ polarizations. All cobalt environments feature a traditional square planar crystal field, placing the $d_{xy}$ orbital between the $d_{x^2-y^2}$ and $d_{xz}/d_{yz}$ orbitals. The resulting orbital configuration is inconsistent with the experimental measurement of the Co $L_3$-edge across both polarizations.}
    \label{fig:ED_fails_suppl}
\end{figure*}

\begin{figure*}[h!]
    \centering
    \includegraphics[width = 85mm]{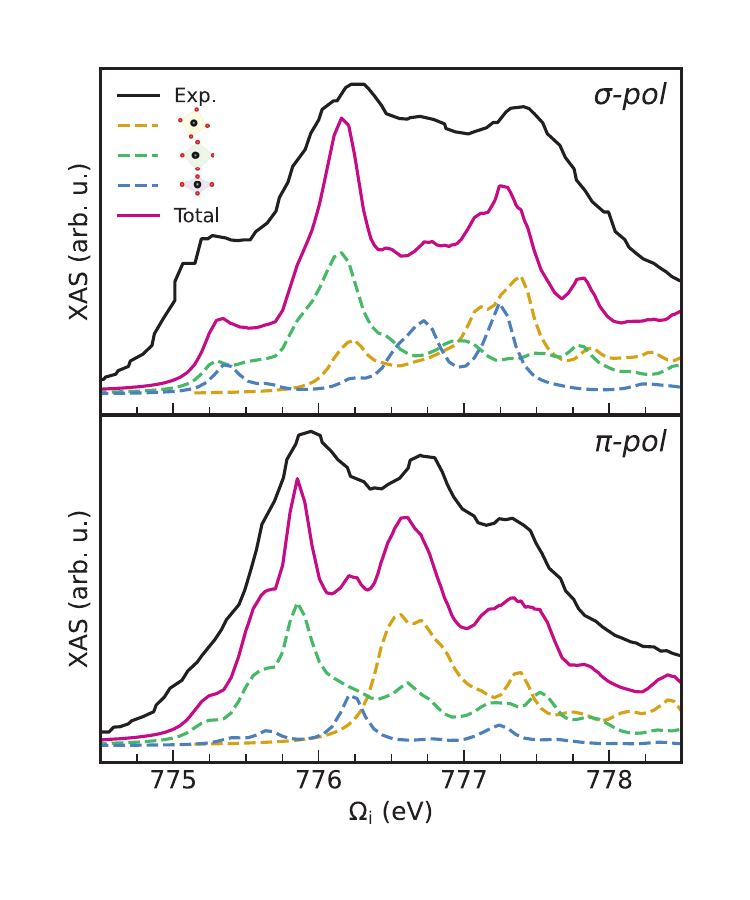}
    \caption{\textbf{Multiplet calculations with a Ca-hybridized square planar crystal field.} Similar to Supplementary Figure S1, individual contributions of each cobalt environment to the overall theoretical result are depicted in colored dashed lines for $\sigma$ and $\pi$ polarizations. Strong hybridization between the Ca 4s, which is now included in the cluster, and the Co $d_{xy}$ orbital lead to a re-arrangement of the crystal field splitting, resulting in appropriate speak splitting across both polarizations of the Co $L_3$-edge.}
    \label{fig:ED_indvidual}
\end{figure*}

% Replaced with Fig 4e
% \begin{figure}[h!]
%     \centering
%     \includegraphics{example-image-a}
%     \caption{Mode in \CCO and comparison with \CCOB.}
%     \label{fig:CCOcompCCOB}
% \end{figure}

\begin{figure*}[h!]
    \centering
    \includegraphics[width = 170mm]{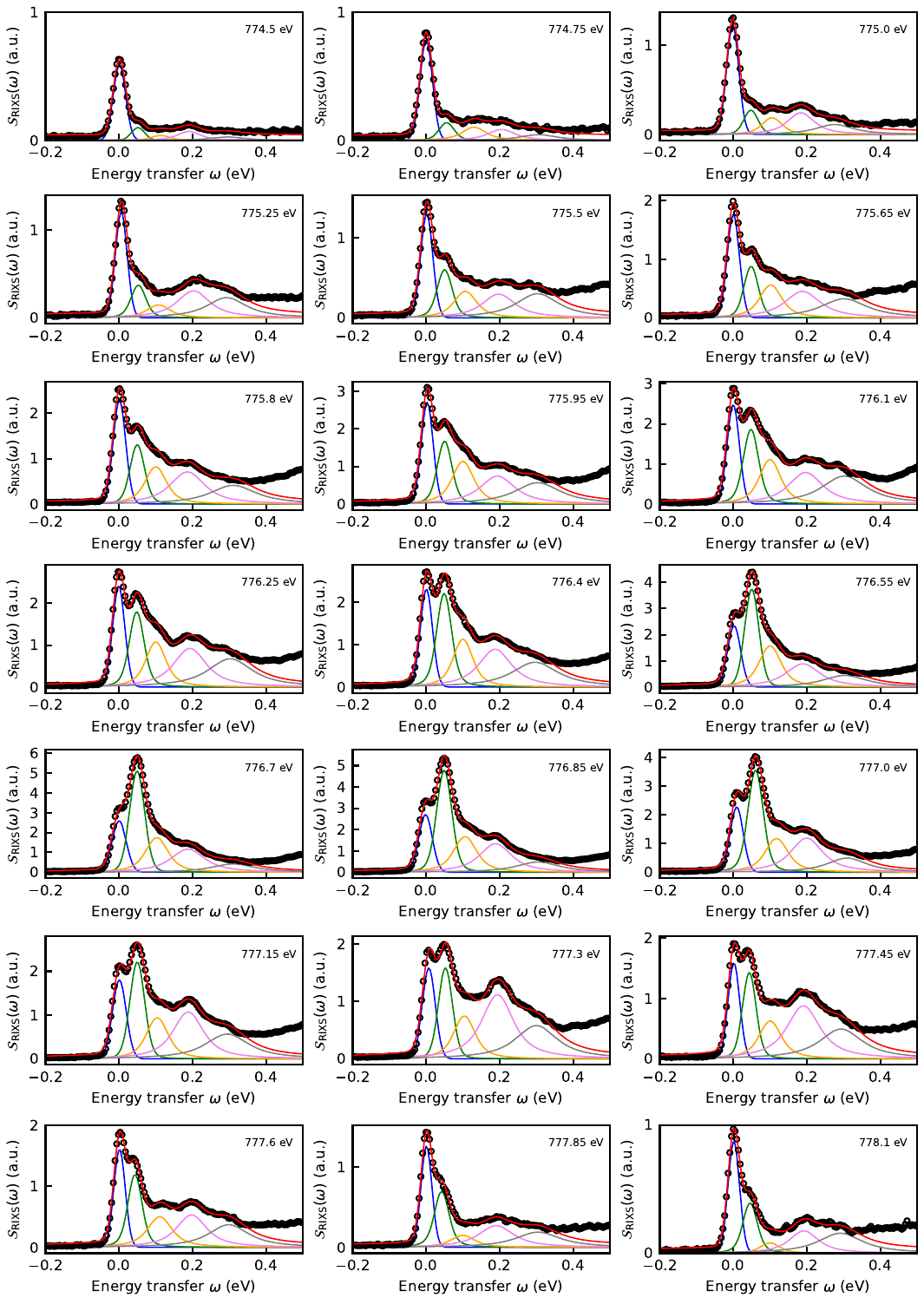}
    \caption{\textbf{Fits of the the incident energy map taken with $\pi$ polarization at an incident angle of $\theta_\mathrm{in} =30^\circ$.} }
    \label{fig:incmapsfits_LH}
\end{figure*}

\begin{figure*}[h!]
    \centering
    \includegraphics[width = 170mm]{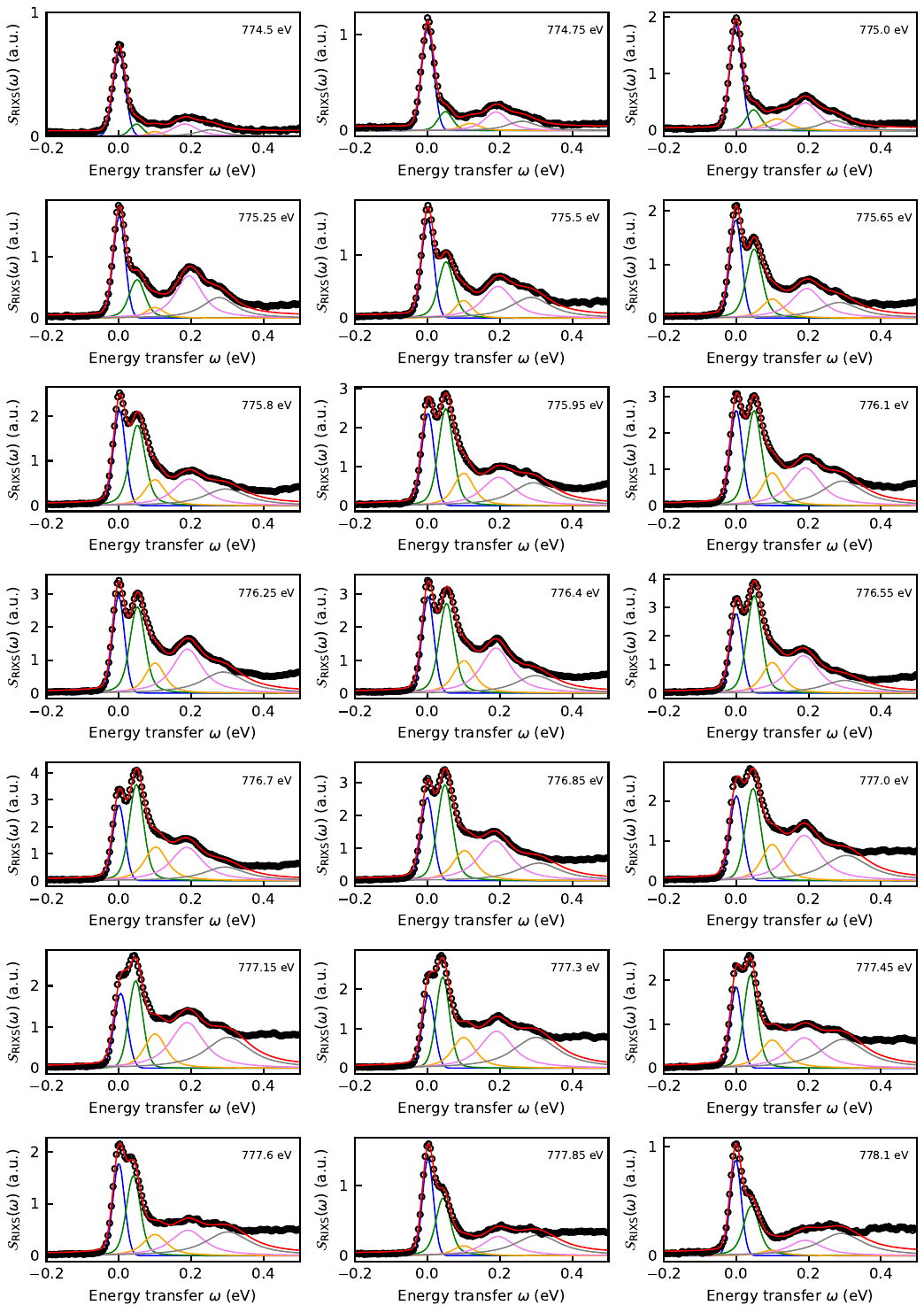}
    \caption{\textbf{Fits of the the incident energy map taken with $\sigma$ polarization at an incident angle of $\theta_\mathrm{in} =30^\circ$.} }
    \label{fig:incmapsfits_LV}
\end{figure*}

\begin{figure*}[h!]
    \centering
    \includegraphics[width = 170mm]{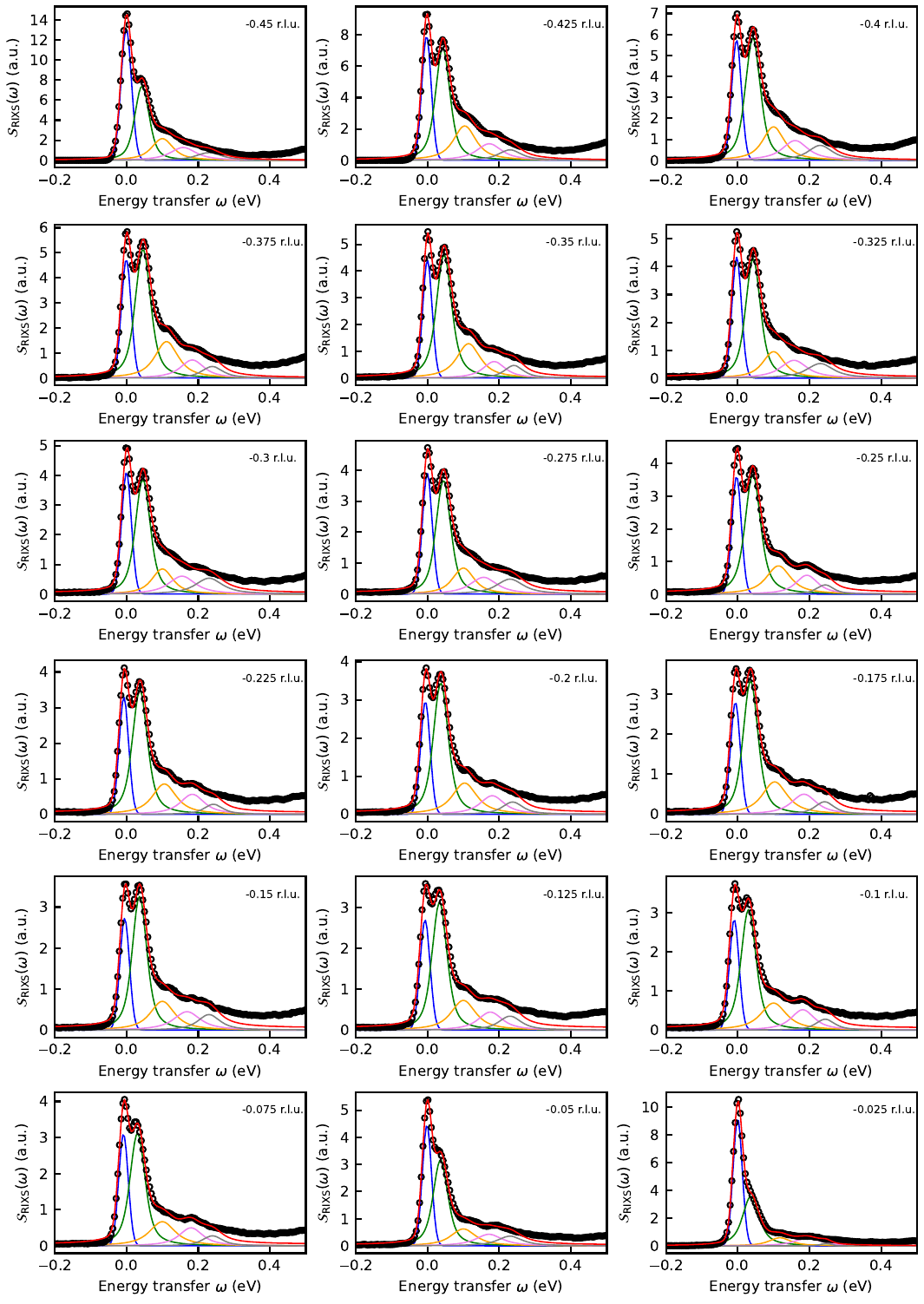}
    \caption{\textbf{Momentum dependence taken along the $(h,0)-$direction of the pseudo-cubic unit cell of CaCoO$_\mathbf{2}$.} }
    \label{fig:qmapsfits_h}
\end{figure*}

\begin{figure*}[h!]
    \centering
    \includegraphics[width = 170mm]{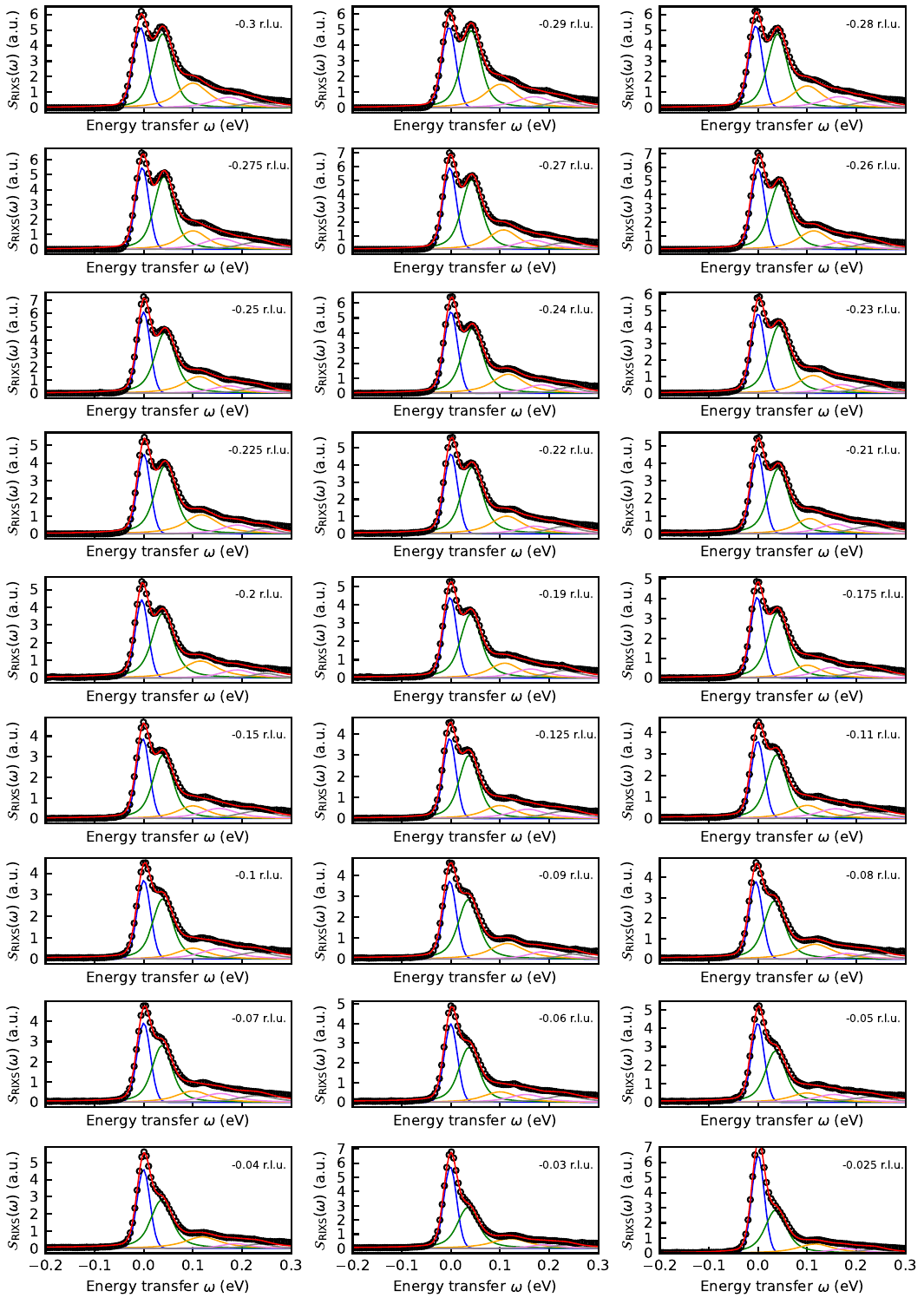}
    \caption{\textbf{Momentum dependence taken along the $(h,k)-$direction of the pseudo-cubic unit cell of CaCoO$_\mathbf{2}$.} }
    \label{fig:qmapsfits_hk}
\end{figure*}

\begin{figure*}[h!]
    \centering
    \includegraphics[width = 170mm]{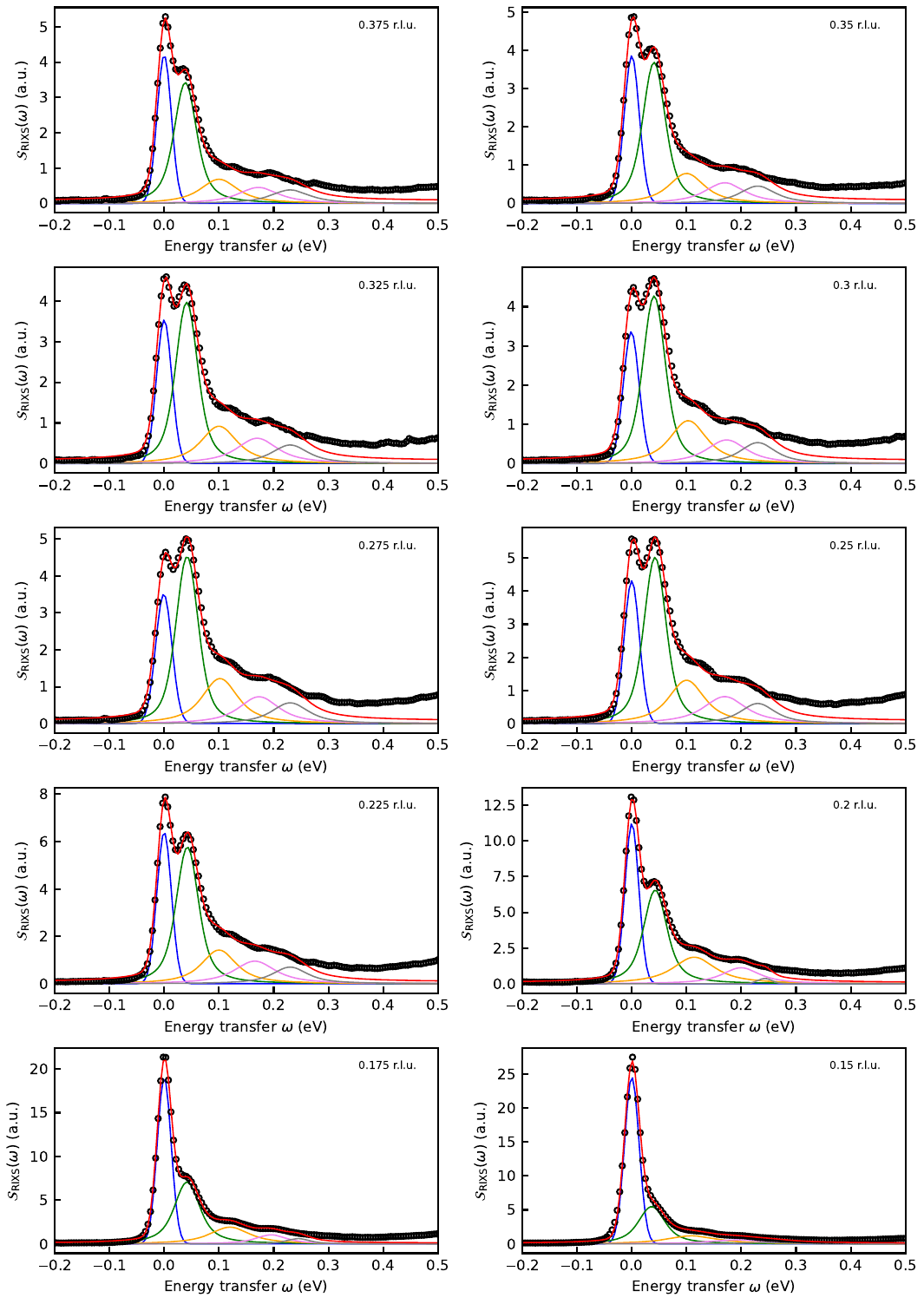}
    \caption{\textbf{Momentum dependence taken along the $(-0.05,-0.05,l)-$direction of the pseudo-cubic unit cell of CaCoO$_\mathbf{2}$.} }
    \label{fig:qmapsfits_l}
\end{figure*}

\end{document}